\documentclass[aps,prl,floatfix,superscriptaddress,twocolumn,longbibliography]{revtex4-2}

\usepackage{graphicx}
\usepackage{dcolumn}
\usepackage{bm}
\usepackage{subfigure}  
\usepackage{amsmath,amssymb}
\def\rb{\doteq}

\usepackage[title]{appendix} 

\usepackage{exscale}
\usepackage{multirow} 
\usepackage{booktabs} 

\usepackage{xcolor}
\usepackage{hyperref}

\newcommand{\ket}[1]{|#1\rangle}
\newcommand{\bra}[1]{\langle#1|}

\definecolor{darkred}{rgb}{0.90,0.2,0.2}

\begin{document}


\title{Onset of Quantum Chaos and Ergodicity in Spin Systems with\\ Highly Degenerate Hilbert Spaces}

\author{Mahmoud Abdelshafy}
\affiliation{Department of Physics, The Pennsylvania State University, University Park, Pennsylvania 16802, USA}

\author{Rubem Mondaini}
\affiliation{Department of Physics, University of Houston, Houston, Texas 77204, USA}
\affiliation{Texas Center for Superconductivity, University of Houston, Houston, Texas 77204, USA}

\author{Marcos Rigol}
\affiliation{Department of Physics, The Pennsylvania State University, University Park, Pennsylvania 16802, USA}


\begin{abstract}
We show that in systems with highly degenerate energy spectra, such as the 2D transverse-field Ising model (2DTFIM) in the strong-field limit, quantum chaos can emerge in finite systems for arbitrary small perturbations. In this regime, the presence of extensive quasiconserved quantities can prevent finite systems from becoming ergodic. We study the ensuing crossover to ergodicity in a family of models that includes the 2DTFIM, in which the onset of ergodic behavior exhibits universality and occurs for perturbation strengths that decrease polynomially with increasing system size. We discuss the behaviors of quantum chaos indicators, such as level spacing statistics and bipartite entanglement, and of the fidelity susceptibilities and spectral functions across the crossover.
\end{abstract}

\maketitle
\textit{Introduction.} The onsets of quantum chaos and ergodicity in isolated clean~\cite{D'Alessio_review_16, Mori_review_18} and disordered~\cite{nandkishore_huse_15, abanin_altman_19, lev_review_24} many-body quantum systems have attracted much attention in the last two decades. A many-body system is said to exhibit quantum chaos when the statistics of the energy spectrum are random-matrix-like, and ergodicity when the eigenstate thermalization hypothesis describes the matrix elements of observables in the energy eigenstates (which guarantees thermalization)~\cite{deutsch_91, Srednicki_94, Rigol_08}. Quantum chaos and ergodicity generally come together~\cite{Santos_10_a,*Santos_10_b}. In clean many-body systems (our interest here), a recurring question has been how the crossover between integrability and quantum chaos and ergodicity occurs in finite systems and how it changes with increasing system size~\cite{rabson_narozhny_04, santos_2004, Santos_10_a,*Santos_10_b, santos_14, Modak_14_a, Modak_14_b, Mondaini_16, Mondaini_17, brenes_leblond_20, Pandey_20, santos_bernal_20, brenes_goold_20, LeBlond_21, schagrin_pozsgay_21, Bulchandani_22, orlov_tiutiakina_23, kim_polkovnikov_24}, as well as how it affects thermalization~\cite{rigol_09a,*rigol_09b}. 

Remarkable findings in clean systems include that of a single integrability-breaking impurity at the center of an integrable interacting chain results in quantum chaos and ergodicity in the thermodynamic limit~\cite{santos_2004, santos_14, brenes_leblond_20, Pandey_20, santos_bernal_20, brenes_goold_20}, while energy transport remains ballistic~\cite{brenes_mascarenas_18}, and thermalization occurs to the thermal prediction at integrability~\cite{brenes_leblond_20} in a time scale that increases with increasing system size~\cite{brenes_leblond_20, Pandey_20}. For extensive perturbations, it was found that the breakdown of integrability occurs for exponentially small (in the system size) perturbation strengths and that it precedes the onset of quantum chaos and ergodicity~\cite{Pandey_20, LeBlond_21, Bulchandani_22, kim_polkovnikov_24}. The latter also occurs for perturbation strengths that vanish in the thermodynamic limit~\cite{LeBlond_21, kim_polkovnikov_24} (as advanced in Refs.~\cite{Santos_10_a,*Santos_10_b, rigol_09a,*rigol_09b}), but in finite systems there always exists a regime in which there is neither integrability nor quantum chaos and ergodicity.

In this Letter we study a family of models that includes the 2D transverse-field Ising model (2DTFIM) in the strong-field limit, which allows us to show that for unperturbed models with highly degenerate spectra the onset of quantum chaos in finite systems can occur for any nonzero perturbation strength. Furthermore, we show that the presence of extensive quasiconserved quantities (the total magnetization in our case) can prevent finite systems from becoming ergodic for perturbation strengths that decrease polynomially with increasing system size. Our results are in stark contrast to those in the weak-field limit of the 2DTFIM, in which the occurrence of Hilbert space fragmentation has attracted attention recently~\cite{Yoshinaga_22, Balducci_22, Hart_22}. The onset of quantum chaos and ergodicity with increasing system size in this limit was studied in Refs.~\cite{Mondaini_16, Mondaini_17}, and quantum quenches in the ferromagnetic phase revealed a lack of thermalization~\cite{blass_rieger_16}. In the strong-field limit, the breakdown of eigenstate thermalization in finite systems was discussed in Refs.~\cite{Mondaini_16, Mondaini_17}.

\textit{Models and calculations.} Motivated by the crossover between ergodic and nonergodic behavior that occurs in the strong-field limit of the spin-$\frac{1}{2}$ 2DTFIM,
\begin{equation}
    \hat H_\text{2DTFIM} \rb \sum_{\bf i}\sigma_{\bf i}^z + J \sum_{\langle {\bf i},{\bf j}\rangle}\sigma_{\bf i}^x \sigma_{\bf j}^x,
    \label{eq:hamiltonian2D}
\end{equation}
where $\langle {\bf i},{\bf j}\rangle$ stands for nearest-neighbor sites ${\bf i}$ and ${\bf j}$, our unperturbed model with a highly degenerate spectrum will be that of spins-$\frac{1}{2}$ in a magnetic field, $\hat H_0 \rb \sum_{\bf i}\sigma_{\bf i}^z$ ($\sigma^{x,z}$~are the $x$ and $z$ Pauli matrices). The eigenenergies of $\hat H_0$ equal the total magnetization, $S_z=\langle \sum_{\bf i}\sigma_{\bf i}^z \rangle$, whose conservation is broken by the interactions $\sum_{\langle {\bf i},{\bf j}\rangle} \sigma_{\bf i}^x \sigma_{\bf j}^x=\sum_{\langle {\bf i},{\bf j}\rangle}(\sigma_{\bf i}^+\sigma_{\bf j}^- + \sigma_{\bf i}^-\sigma_{\bf j}^+ +\sigma_{\bf i}^+ \sigma_{\bf j}^+ + \sigma_{\bf i}^- \sigma_{\bf j}^-)$. In two and higher dimensions, the models obtained by adding $\sum_{\langle {\bf i},{\bf j}\rangle} \sigma_{\bf i}^x \sigma_{\bf j}^x$, or only the magnetization-breaking subterms $\sum_{\langle {\bf i},{\bf j}\rangle}(\sigma_{\bf i}^+ \sigma_{\bf j}^+ + \sigma_{\bf i}^- \sigma_{\bf j}^-)$, to $\hat H_0$ are nonintegrable. In 1D, those models are integrable, and next-nearest-neighbor terms can be added to break integrability.

To study the crossover to quantum chaos and ergodicity with increasing system size when $\hat H_0$ is perturbed, we consider 1D models of the form:
\begin{eqnarray}
 &&\hat H_\text{1D} \! \rb \! \sum_{i}\sigma_{i}^z + 4J\sum_{i} V_i, \nonumber \\
 && V_i \! = \! \sigma_{ i}^+\sigma_{i+1}^++\sigma_{i}^-\sigma_{i+1}^-+\sigma_{i}^+\sigma_{i+2}^++\sigma_{i}^-\sigma_{i+2}^-\,,
 \label{eq:hamiltonian1D}
\end{eqnarray}
in chains with periodic boundary conditions. We carry out full exact diagonalization calculations of chains with up to $L=22$ sites after taking into account the symmetries (discrete translations and $Z_2$ in the $x$ direction). Due to the higher computational cost, for $L=22$ we only diagonalize one quasimomentum sector ($k=6\pi/11$).

We compute the average $r_\text{ave} \equiv \overline{r_n}$ of the ratio of the smallest to the largest consecutive level spacings $r_n=\text{min}(\delta_{n},\delta_{n+1})/\text{max}(\delta_{n},\delta_{n+1})$, where $\delta_{n}=E_{n+1}-E_{n}$ and $E_{n}$ is the $n$-th eigenenergy~\cite{Oganesyan_07}. We also compute the normalized average $s_\text{ave}\equiv\overline{S^{(n)}_A}/(\frac{L}{2}\ln2)$ of the bipartite entanglement entropy of energy eigenstates $\ket{\psi_n}$, $S^{(n)}_{A}=-\text{Tr}(\hat\rho^{(n)}_{A}\ln\hat\rho^{(n)}_{A})$~\cite{Bianchi_22, patil_23, rafal_24, yauk_24}. To calculate the reduced density matrix $\hat\rho^{(n)}_{A}$, we trace out the complement $B$ of subsystem $A$, $\hat\rho^{(n)}_{A}=\text{Tr}_{B}(\ket{\psi_n}\bra{\psi_n})$, and focus on the case where $A$ and $B$ are composed of $L/2$ contiguous sites. The averages $r_\text{ave}$ and $s_\text{ave}$ are calculated in the central 20\% of the energy spectrum. 

We also study the diagonal and off-diagonal matrix elements of various observables $\hat{O}$ in the energy eigenstates. Associated with the off-diagonal matrix elements, we study the low-frequency behavior of the average spectral function $F^{O}_\text{ave}\!\equiv\!\overline{|f_{n}^{O}(\omega)|^{2}}$, with
\begin{equation}
 |f_{n}^{O}(\omega)|^{2} = L\sum_{m\ne n}|\bra{n}\hat{O}\ket{m}|^{2}\delta(\omega-\omega_{nm})\ ,
 \label{eq:Spectral_function}
\end{equation}
where $\omega_{nm}\equiv E_{n}-E_{m}$ and $\delta(\omega)$ is the Dirac-delta function~\footnote{We regularize the $\delta$ function using the Gaussian function $\exp(-\frac{x^2}{2\eta^2})\!/\!\left(\sqrt{2\pi}\eta\right)$, where $\eta\!=\!\omega_{\rm min}$ denotes a cutoff frequency with $\omega_{\rm min}\!=\!\min_n(E_{n+1}\!-\!E_{n})$. In the SM~\cite{SM}, we show that similar results are obtained using a different regularization.}. We further study the typical fidelity susceptibility~\cite{LeBlond_21, Dries_21}, $\chi^O_\text{typ}\!\!=\!\exp(\overline{\ln\left[\chi^O_{n}\right]})$, where
\begin{equation}
 \chi^O_{n} = L\sum_{m\ne n}\frac{|\bra{n}\hat{O}\ket{m}|^{2}}{\left(E_{n}-E_{m}\right)^{2}}.
 \label{eq:susceptibility}
\end{equation}
The results for $F^{O}_\text{ave}$ and $\chi^O_\text{typ}$ are obtained averaging over the central 20\% and the entire spectrum, respectively.

\begin{figure}[!t]
    \includegraphics[width=0.985\columnwidth]{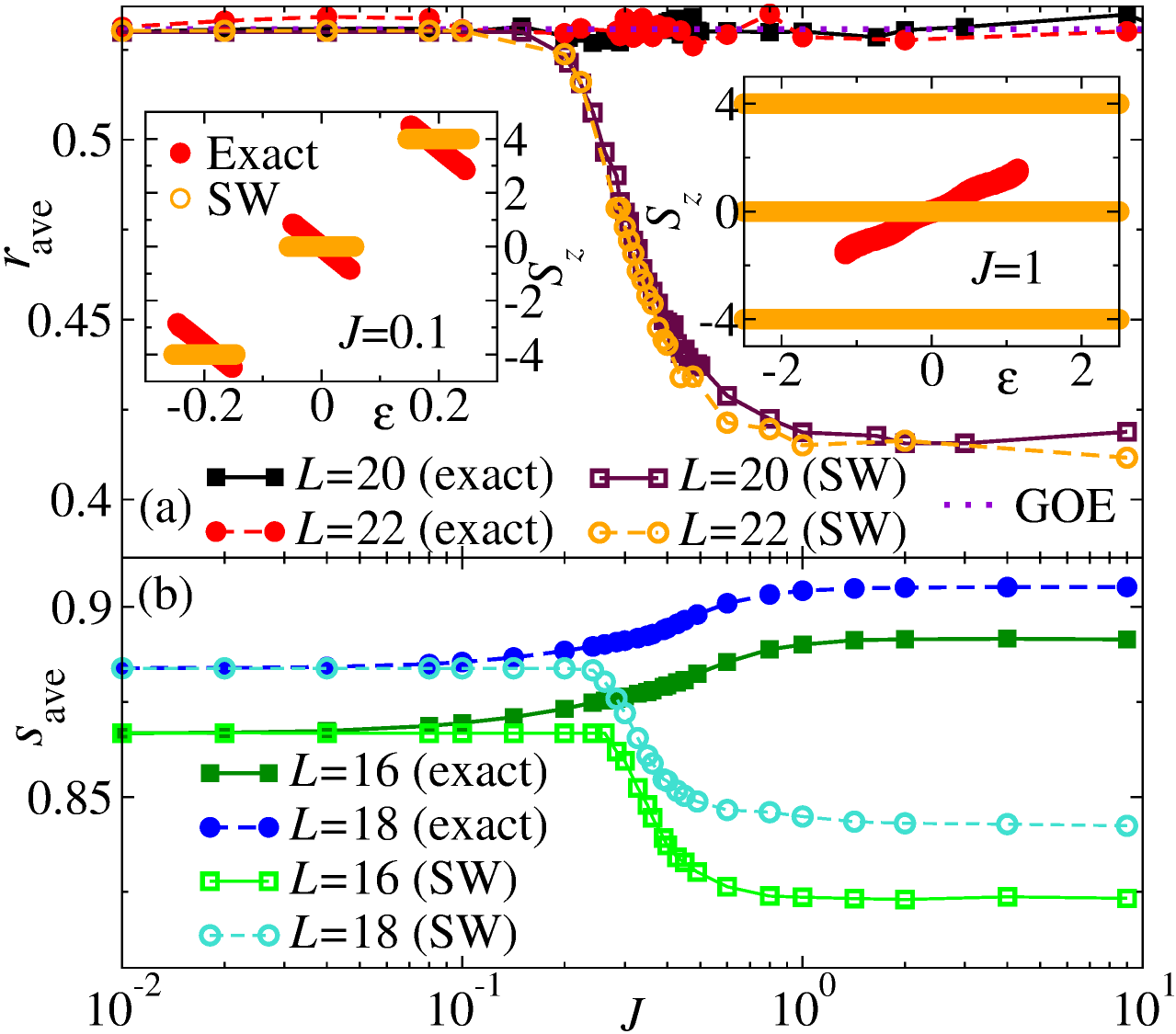}
    \vspace{-0.1cm}
    \caption{(a) Average ratio of consecutive level spacings $r_\text{ave}$ vs~the perturbation strength $J$ for $L=20$ and $22$. We average over all quasimomentum $k$ sectors with $k\ne0,\pi$ for $L=20$, $k=6\pi/11$ for $L=22$, and the two $Z_{2}$ (in the $x$ direction) symmetry sectors. The horizontal dotted line shows $r_\text{ave}$ for the Gaussian orthogonal ensemble (GOE)~\cite{Atas_13}. The left (right) inset shows the magnetization $S_{z}$ in the energy eigenstates vs~their energy density $\varepsilon$ at $J=0.1$ ($J=1$) for $L=22$, $k=6\pi/11$, and $Z_{2}=-1$. (b) Normalized average bipartite entanglement entropy $s_\text{ave}$ in the sector with quasimomentum $k=\pi/2$, $Z_{2}=1$ for $L=16$, and $k=4\pi/9$, $Z_{2}=-1$ for $L=18$. Results are reported for $\hat H_\text{1D}$ (solid symbols), and $\hat H_\text{1DSW}$ (open symbols), and were obtained averaging over the central 20\% of the spectrum of each symmetry subspace.}
    \label{fig:r_S}
\end{figure}

\textit{Results.} In Fig.~\ref{fig:r_S}(a) we plot $r_\text{ave}$ versus $J$ (solid symbols) over three decades of values of $J$. Notably, except for small dips and enhanced fluctuations for $ 0.2 \lesssim J \lesssim 0.6$, $r_\text{ave}\approx 0.53$ as predicted for the Gaussian orthogonal ensemble~\cite{Atas_13}. This indicates that the model is quantum chaotic from arbitrarily small to arbitrarily large values of $J$. The fact that something changes in the nature of the energy eigenstates in between becomes apparent only in other quantities, such as the normalized average eigenstate entanglement entropy $s_\text{ave}$ [see Fig.~\ref{fig:r_S}(b)]. $s_\text{ave}$ is constant and its magnitude~\footnote{In the thermodynamic limit, $s_\text{ave}=1$ for quantum-chaotic models and $s_\text{ave}\approx 0.55$ for integrable ones~\cite{Bianchi_22}.} indicates quantum chaotic behavior both at small and large values of $J$, but exhibits a crossover for $ 0.2 \lesssim J \lesssim 0.6$ (in the system sizes shown) in which it increases with increasing $J$.

As $J\rightarrow0$, we can understand our results by computing the effect of the perturbation on the degenerate subspaces, e.g., via a Schrieffer-Wolff (SW) transformation~\cite{Schrieffer_66, BRAVYI_11}. For our model, to order $J^2$, the SW Hamiltonian reads [see Supplemental Material (SM)~\cite{SM}]
\begin{equation}
\hat H_\text{1DSW}\! \rb\! \left(1+8J^{2}\right)\!\! \sum_{i} \! \sigma_{i}^z + 4J^{2}\sum_{i} \! \sigma_{i}^z \!\!\!\!\!\!\!  \sum_{j_i,k_i:k_i>j_i} \!\!\!\!\!\!\!\!\! \left( \sigma_{j_i}^+ \sigma_{k_i}^- + \sigma_{j_i}^ - \sigma_{k_i}^+ \right)\!,
\label{eq:SW1D} 
\end{equation}
where ${j_i,k_i}\in \left\{{i-2,i-1,i+1,i+2}\right\}$. The magnetization, which commutes with $\hat H_\text{1DSW}$, is quasiconserved for small $J$ in our system.

\begin{figure}[!t]
    \includegraphics[width=0.985\columnwidth]{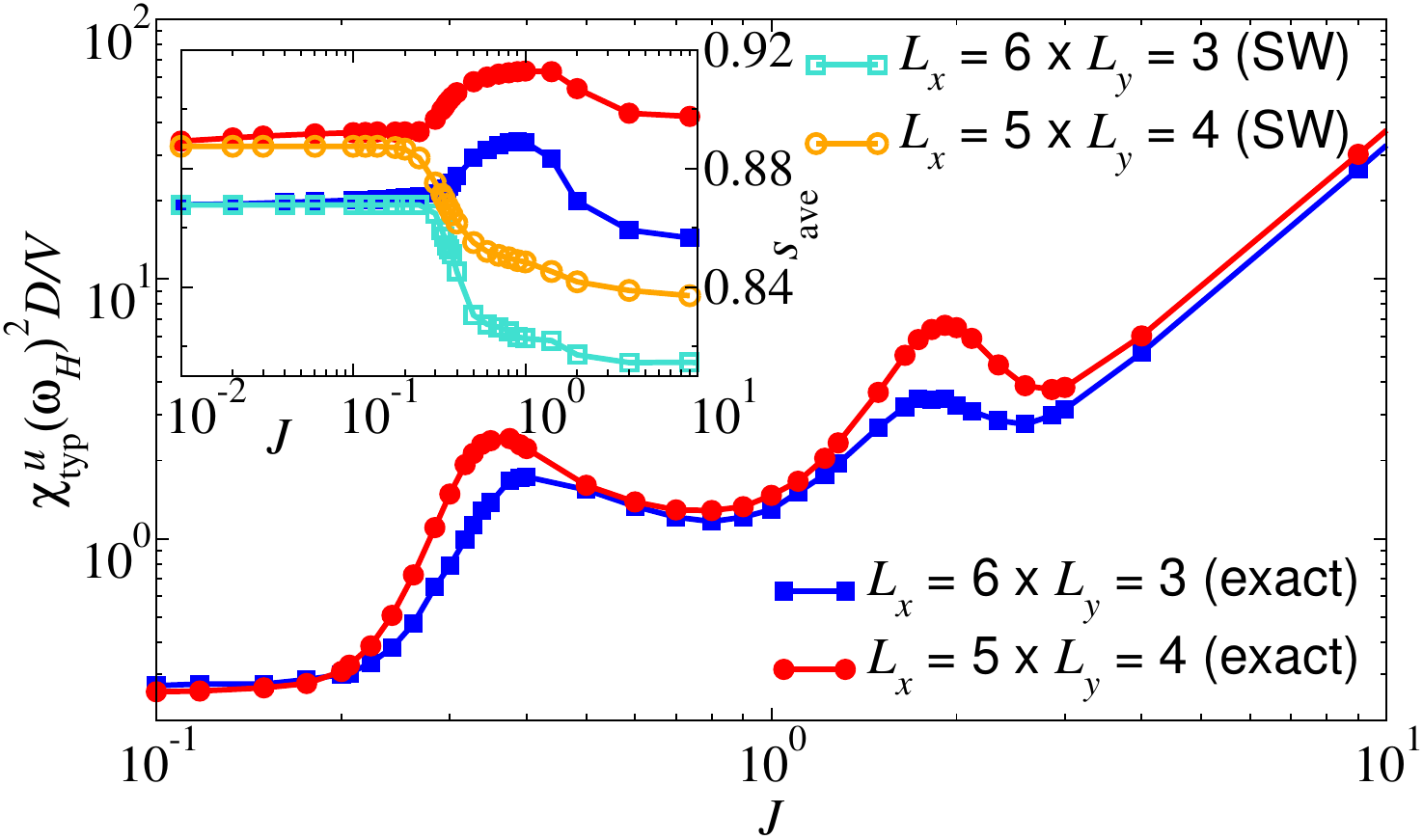}
    \vspace{-0.1cm}
    \caption{Rescaled typical fidelity susceptibility $\chi^u_\text{typ}$ vs the perturbation strength $J$ for 2D lattices with $L_x=6,\,L_y=3$ and $L_x=5,\,L_y=4$. We report results obtained in the quasimomentum $k=(0,0)$ sector, averaged over all states in the $Z_{2}$, $M_{x}$, and $M_{y}$ subsectors ($\hat M_{x}$ and $\hat M_{y}$ stand for mirror symmetry in $x$ and $y$, respectively). Inset: Normalized average bipartite entanglement entropy $s_\text{ave}$ in the $k=(0,0)$ subsector with $Z_{2}=-1$, $M_{x}=-1$, and $M_{y}=-1$ ($Z_{2}=1$, $M_{x}=1$ and $M_{y}=1$) for $L_x=6,\,L_y=3$ ($L_x=5,\,L_y=4$). Results are reported for $\hat H_\text{2DTFIM}$ (solid symbols), and $\hat H_\text{2DPT}$ (open symbols), and were obtained averaging over the central 20\% of the spectrum of each symmetry subspace.}
    \label{fig:2D_TFIM}
\end{figure}

In Figs.~\ref{fig:r_S}(a) and~\ref{fig:r_S}(b), we plot $r_\text{ave}$ and $s_\text{ave}$, respectively, versus $J$ for $\hat H_\text{1DSW}$ (open symbols). The results are indistinguishable from those for $\hat H_\text{1D}$ (the ``exact'' results) for small $J$. Their agreement makes apparent that quantum chaos emerges in $\hat H_\text{1D}$ for arbitrarily small $J$ because $\hat H_\text{1DSW}$ is already quantum chaotic to lowest order in perturbation theory. This is something that can occur for a wide range of systems and, in fact, also occurs for the 2DTFIM (see Fig.~\ref{fig:2D_TFIM}). When $J\gtrsim 0.1$, the exact and SW results differ from each other because the magnetization is not (is) conserved in $\hat H_\text{1D}$ ($\hat H_\text{1DSW}$). The insets in Fig.~\ref{fig:r_S}(a) show the magnetization $S_z$ in the energy eigenstates versus their energy density $\varepsilon=E_n/L$ just before ($J=0.1$) and after ($J=1$) the results for $\hat H_\text{1D}$ and $\hat H_\text{1DSW}$ depart from each other in the main panels. We report results for the central $\sim80\%$ of the energy spectrum in a chain with $L=22$ sites, which include the $S_z=0,\,\pm4$ magnetization sectors for $J=0.1$. For $\hat H_\text{1DSW}$, all that happens with increasing $J$ is that the eigenenergies of different magnetization sectors overlap with each other (there is no change in the statistics of the level spacings in each magnetization sector). Since there is no level repulsion, this results in a decrease of $r_\text{ave}$ seen in Fig.~\ref{fig:r_S}(a). In the eigenstates of $\hat H_\text{1D}$, there is ``hybridization'' between different magnetization sectors (note that the magnetization is not constant), resulting in level repulsion as $J$ increases and the sectors overlap.

\begin{figure}[!t]
    \includegraphics[width=\columnwidth]{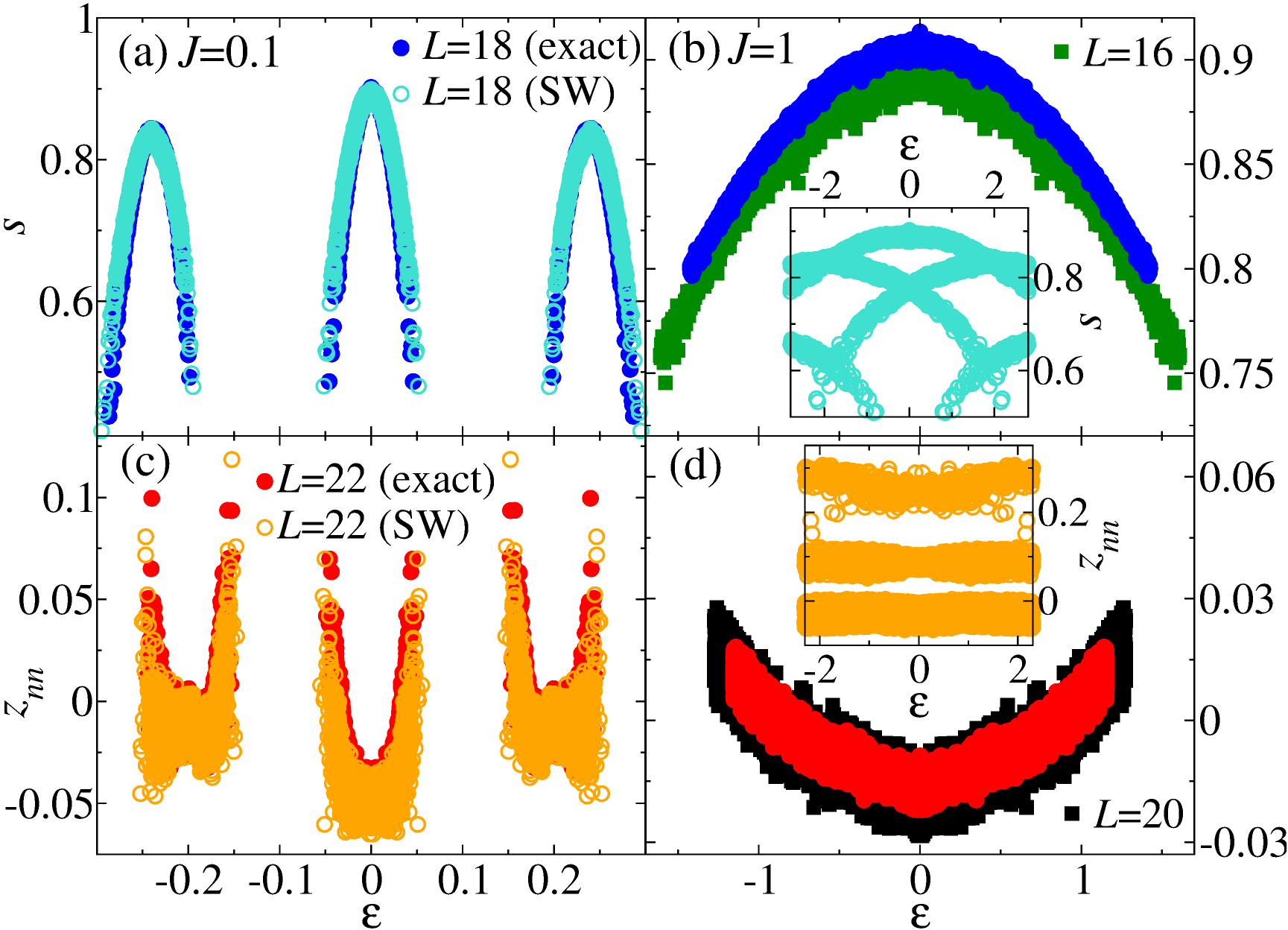}
    \vspace{-0.5cm}
    \caption{(a),(b) Normalized bipartite entanglement entropy of energy eigenstates $s$ vs $\varepsilon$ in the central $\sim 80\%$ of the energy spectrum. (a) Exact and SW results for $J=0.1$ and $L=18$ (b) Exact (main panel, $L=16$ and 18) and SW (inset, $L=18$) results for $J=1$. The results for $L=18$ ($L=16$) are from the sector with $k=4\pi/9$, $Z_{2}=-1$ ($k=\pi/2$, $Z_{2}=1$). (c),(d) Same as (a),(b) but for the nearest-neighbor $z$-$z$ correlations $z_\text{nn}$ in the energy eigenstates of a chain with $L=22$ ($L=20$) in the sector with $k=4\pi/9$, $Z_{2}=-1$ ($k=\pi/2$, $Z_{2}=1$).}
    \label{fig:entropy_Snn}
\end{figure}

The smooth behaviors of the eigenstate magnetization versus $\varepsilon$ in the insets in Fig.~\ref{fig:r_S}(a) suggest that eigenstate thermalization occurs together with quantum chaos for small (large) $J$ within each sector with quasiconserved magnetization (through the entire spectrum). In Figs.~\ref{fig:entropy_Snn}(a) and~\ref{fig:entropy_Snn}(b), we plot the normalized bipartite entanglement entropy of energy eigenstates $s=S^{(n)}_{A}/(\frac{L}{2}\ln2)$ versus $\varepsilon$ for $J=0.1$ and $J=1$, respectively. For $J=0.1$, $s$ in $\hat H_\text{1D}$ is a smooth function of $\varepsilon$ only in each magnetization sector (the ones shown have the same energy density up to a subextensive $1/L$ difference), and it is closely followed by the SW results. For $J=1$, on the other hand, $s$ in $\hat H_\text{1D}$ is a smooth function of $\varepsilon$ throughout the spectrum. At fixed $\varepsilon$, as expected~\cite{Bianchi_22}, $s$ slightly increases (while its eigenstate-to-eigenstate fluctuations decrease) with increasing system size [in Fig.~\ref{fig:entropy_Snn}(b) we show results for two values of $L$]. $s$ in $\hat H_\text{1DSW}$ behaves starkly differently because the results for different magnetization sectors simply overlap with each other at the center of the energy spectrum [inset in Fig.~\ref{fig:entropy_Snn}(b)]. Since the eigenstates of $\hat H_\text{1DSW}$ with $S_z=\pm4$ have lower entanglement entropy than those with $S_z=0$~\cite{Bianchi_22}, as the three sectors overlap with increasing $J$ the average $s_\text{ave}$ decreases as seen in Fig.~\ref{fig:r_S}(b). In contrast, $s_\text{ave}$ for $\hat H_\text{1D}$ increases due to the ``hybridization'' of the magnetization sectors.

In Figs.~\ref{fig:entropy_Snn}(c) and~\ref{fig:entropy_Snn}(d), we plot the eigenstate expectation values of the nearest-neighbor $z$-$z$ correlations $z_\text{nn}=\langle \sum_{i} \sigma_{i}^z \sigma_{i+1}^z \rangle/L$ versus $\varepsilon$ (for $\sim$80\% of the energy spectrum in chains with $L=20$ and 22 sites). The behaviors of $z_\text{nn}$, and the comparison between the exact and the SW results, are qualitatively similar to those of $s$~\footnote{Note that since the chains in the top and bottom panels have different system sizes, the $\varepsilon$ ranges are slightly different in the results}. Our findings for the eigenstate entanglement entropy, the magnetization, and nearest-neighbor $z$-$z$ correlations indicate that at small $J$ there is quantum chaos and eigenstate thermalization within sectors of the energy spectrum in which the magnetization is quasiconserved. The system is not ergodic in that regime. For large $J$, quantum chaos and eigenstate thermalization occur across the energy spectrum, and the system is ergodic.

\begin{figure}[!t]
    \includegraphics[width=0.985\columnwidth]{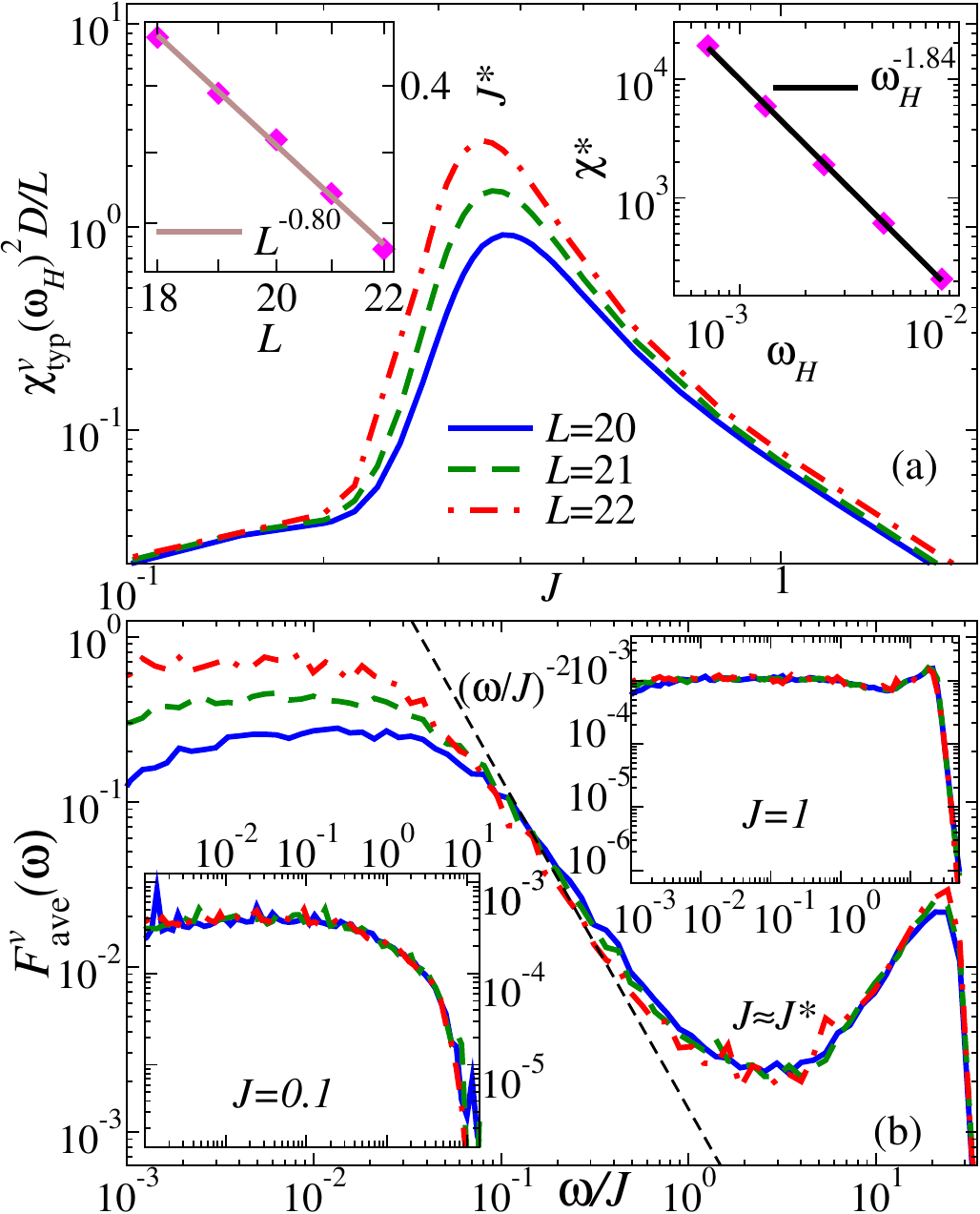}
    \vspace{-0.1cm}
    \caption{(a) Rescaled typical fidelity susceptibility $\chi^v_\text{typ}$ vs the perturbation strength $J$. The left inset shows the position $J^*$ of the maximum of $\chi^v_\text{typ}$ vs the chain size $L$, and the outcome of a fit to $aL^b$ with $a$ and $b$ as fitting parameters. The right inset shows the maximum $\chi^*$ of $\chi^v_\text{typ}$ vs $\omega_H$ (mean level spacing), and the outcome of a polynomial fitting to $a\omega_H^b$. ($J^*$ and $\chi^*$ are computed via a quadratic fit of the data about the maxima.) (b) Spectral function $F^{u}_\text{ave}$ vs $\omega/J$ for $J\!\approx\!J^{*}$ (main panel), $J\!=\!0.1$ (bottom inset), and $J\!=\!1$ (top inset). The straight dashed line in the main panel shows $(\omega/J)^{-2}$ behavior. We show results for $L=20$ and $21$ ($L=22$) computed as a weighted average over the two $Z_{2}$ sectors and the quasimomentum sectors with $k\ne 0,\pi$ ($k=6\pi/11$).}
    \label{fig:chi_spectf}
\end{figure}

Next, we use the typical fidelity susceptibility $\chi^O_\text{typ}$ and the spectral function $F^{O}_\text{ave}$ to study the crossover between these two regimes and the values of $J$ for which it occurs with increasing system size. As observable for these calculations, we take $ v=(\sum_{i} V_i)/L$, where $V_i$ is defined in Eq.~\eqref{eq:hamiltonian1D}. In Fig.~\ref{fig:chi_spectf}(a), we plot the typical fidelity susceptibility $\chi^v_\text{typ}$ versus $J$ in the crossover region. The rescaling used for $\chi^v_\text{typ}$, involving the dimension of the Hilbert space $D$, the system size $L$, and the mean level spacing $\omega_H$, leads to a collapse of the curves for different system sizes when eigenstate thermalization occurs~\cite{LeBlond_21}. The results for three different system sizes in Fig.~\ref{fig:chi_spectf}(a) show that between the two quantum-chaotic regimes (for small and large values of $J$), a peak develops in $\chi^v_\text{typ}$ and it diverges with increasing system size.

As universally found in studies of clean~\cite{LeBlond_21, orlov_tiutiakina_23, kim_polkovnikov_24}, disordered~\cite{LeBlond_21, Dries_21}, and driven~\cite{bhattacharjee_24} systems, the maximum value $\chi^*$ of the susceptibility at the peak exhibits a divergence consistent with $\chi^*\propto \omega^{-2}_H$ [see the right inset in Fig.~\ref{fig:chi_spectf}(a)]. This is as fast as $\chi$ can diverge in finite systems~\cite{Pandey_20}. Furthermore, the position $J^*$ of $\chi^*$ shifts to smaller values as the size of the system increases. The left inset in Fig.~\ref{fig:chi_spectf}(a) shows that the shift is polynomial in the system size ($\propto L^{-0.80}$ within the system sizes considered), as opposed to the exponential shift with system size found in the crossover between integrable interacting and nonintegrable regimes in Ref.~\cite{LeBlond_21}. No faster than a polynomial dependence of the crossover is expected for our system because the support of the fixed magnetization energy bands generated by $\hat H_\text{1DSW}$ is $\propto J^2 L$, and those bands need to overlap for the system to become ergodic. Hence, $(J^*)^2 L = O(1)$ or $J^*\propto 1/\sqrt L$ is the fastest that $J^*$ can decrease with the system size. The faster decrease observed in our system sizes is an expected consequence of finite-size effects. We stress that our results indicate that, in the thermodynamic limit, the system is ergodic for arbitrarily small perturbation strengths.

In Fig.~\ref{fig:chi_spectf}(b), we show the behavior of the spectral function $F^{v}_\text{ave}$ versus $\omega$ for $J=0.1$ (bottom inset), $J\approx J^*$ (main panel), and $J=1$ (top inset). For $J=0.1$ and 1, the spectral function exhibits the low-frequency plateau expected for quantum-chaotic systems~\cite{D'Alessio_review_16}. On the other hand, for $J\approx J^*$, the spectral function at low-frequency diverges with increasing system size. This explains the divergence of $\chi^v_\text{typ}$ seen in Fig.~\ref{fig:chi_spectf}(a) at $J^*$, and indicates that in the thermodynamic limit the thermalization times diverge as $J\rightarrow 0$. More importantly, about $J\approx J^*$, one can see that with increasing system size, the divergence of the spectral function is consistent with being $\propto (\omega/J)^{-2}$ (see the dashed line in the plot). This $\omega$ dependence is the one expected from Fermi's golden rule and was also observed and discussed in the context of the integrability to nonintegrability crossover in spin chains~\cite{LeBlond_21}.

To conclude, we briefly discuss exemplary results~\footnote{A full scaling analysis such as the one done for chains cannot be carried out in two dimensions because of computational limitations due to the rapid increase of the number of lattice sites with the linear dimension, which are compounded with strong finite-size effects.} for the fidelity susceptibility and the bipartite entanglement entropy of the 2DTFIM [see Eq.~\eqref{eq:hamiltonian2D}] in periodic lattices with $L_x$ ($L_y$) sites in the $x$ ($y$) direction. To order $J$, the perturbed Hamiltonian for this model only contains the magnetization preserving part of $J \sum_{\langle {\bf i},{\bf j}\rangle}\sigma_{\bf i}^x \sigma_{\bf j}^x$: $\hat H_\text{2DPT}\rb \sum_{\bf i} \sigma_{\bf i}^z + J \sum_{\langle {\bf i},{\bf j}\rangle}( \sigma_{\bf i}^+  \sigma_{\bf j}^-+ \sigma_{\bf i}^-  \sigma_{\bf j}^+)$, and is nonintegrable. We study $\chi^u_\text{typ}$ for $ u\!=\![\sum_{\langle {\bf i},{\bf j}\rangle}( \sigma_{\bf i}^+ \sigma_{\bf j}^+\!+\! \sigma_{\bf i}^- \sigma_{\bf j}^-)]/V$, where $V=L_x\times L_y$ is the number of lattice sites. In Fig.~\ref{fig:2D_TFIM}, we plot $\chi^u_\text{typ}$ versus $J$ for two lattice sizes. Two crossovers are highlighted by the susceptibility peaks; the first one is the one that parallels the crossover studied before in chains [see Fig.~\ref{fig:chi_spectf}(b)], and the second one is the crossover away from ergodicity when approaching the classical Ising limit. In the SM~\cite{SM}, we show that the corresponding spectral functions behave qualitatively as those in Fig.~\ref{fig:chi_spectf}(b). The results for $s_\text{ave}$ (shown in the inset) also indicate the presence of these crossovers, with the ergodic regime exhibiting the maximal entanglement entropy. For small values of $J$, as in 1D, the results for $s_\text{ave}$ in $\hat H_\text{2DPT}$ closely follow the exact ones.

\textit{Summary.} We showed that in systems with highly degenerate energy spectra, which usually occur in high-field and strong-interaction limits of models of interest in different areas of physics, quantum chaos can emerge in finite systems for arbitrary small perturbations. This might appear counterintuitive as one does not expect systems to be ergodic in such regimes~\cite{Mondaini_16, Mondaini_17, kollath07, roux09, *roux10, rigol10, biroli10}. We find that the lack of ergodicity can be a finite-size effect of the presence of quasiconserved quantities that disappear in the thermodynamic limit. In the latter limit, thermalization can ultimately occur regardless of how strong are the field or interactions, but the stronger they are, the longer it will take the system to thermalize (as shown by our diverging low-frequency spectral functions). We also showed that in finite systems, the crossover to ergodicity is marked by a universal divergence of the typical fidelity susceptibility and an increase of the eigenstate entanglement. We considered here the case in which all degeneracies disappear at the lowest order in perturbation theory, yet what happens when that is not the case is an interesting open question we plan to explore next, along with the possibility of the perturbative regime being integrable.

\acknowledgments
This work was supported by the National Science Foundation (NSF) Grant No.~PHY-2309146 (M.A.~and M.R.) and the T$_{\rm c}$SUH Welch Professorship Award. (R.M.). M.R.~is grateful to David Huse and Sarang Gopalakrishnan for stimulating discussions.

\bibliography{Reference}

\setcounter{figure}{0}
\setcounter{equation}{0}
\setcounter{table}{0}

\renewcommand{\thetable}{S\arabic{table}}
\renewcommand{\thefigure}{S\arabic{figure}}
\renewcommand{\theequation}{S\arabic{equation}}

\renewcommand{\thesection}{S\arabic{section}}

\onecolumngrid

\vspace*{0.35cm}

\begin{center}

{\large \bf Supplemental Material:\\
Onset of Quantum Chaos and Ergodicity in Spin Systems with\\ Highly Degenerate Hilbert Spaces}\\

\vspace{0.3cm}

Mahmoud Abdelshafy$^{1}$, Rubem Mondaini$^{2,3}$, Marcos Rigol$^{1}$

$^{1}${\it Department of Physics, The Pennsylvania State University, University Park, Pennsylvania 16802, USA}\\
$^{2}${\it Department of Physics, University of Houston, Houston, Texas 77004, USA}\\
$^{3}${\it Texas Center for Superconductivity, University of Houston, Houston, Texas 77204, USA}

\end{center}

\vspace{0.6cm}

\twocolumngrid

\label{pagesupp}

\section{Schrieffer-Wolff (SW) Hamiltonian\label{sec_S1}}

The Schrieffer-Wolff (SW) Hamiltonian is obtained by applying a unitary transformation to the exact Hamiltonian $\hat H=\hat H_0+g\hat V$, resulting in an effective Hamiltonian that describes the low-energy subspaces,
\begin{equation}
 \hat H_{\rm SW}=e^{\hat S} \hat H e^{-\hat S},
 \label{eq:SWTransformation}
\end{equation}
where the anti-Hermitian operator $\hat S=-\hat S^{\dagger}$ is the generator of the transformation. Using the Baker–Campbell–Hausdorff formula, and imposing the constraint $[\hat S,\hat H_0]=-g\hat V$, results in
\begin{equation}
 \hat H_{\rm SW}=\hat H_0+ 
 \frac{g}{2}[\hat S,\hat V]+O(g^3),
 \label{eq:SW_2ndorder}
\end{equation}
which gives $\hat H^{\rm SW}$ up to second-order in $g$, because $\hat S\propto g$.

In our case, $\hat H_{0} \rb \sum_i \sigma_i^z$, and 
\begin{equation}
\hat V \rb \sum_{i}\left(\sigma_{i}^+\sigma_{i+1}^++ \sigma_{i}^-\sigma_{i+1}^- + \sigma_{i}^+ \sigma_{i+2}^++ \sigma_{i}^-\sigma_{i+2}^-\right).
\end{equation}
The generator $\hat S$ of the transformation for this model is
\begin{equation}
 \hat S \rb \frac{g}{4} \sum_{i}\left(\sigma_{i}^+\sigma_{i+1}^+-\sigma_{i}^-\sigma_{i+1}^- + \sigma_{i}^+\sigma_{i+2}^+- \sigma_{i}^-\sigma_{i+2}^-\right),
 \label{eq:S}
\end{equation}
and, taking $g=4J$, one obtains $H_\text{1DSW}$ in Eq.~\eqref{eq:SW1D} in the main text.

\section{Gaussian versus Lorentzian Regularization\label{sec_S2}}

\begin{figure}[!h]
    \includegraphics[width=.985\columnwidth]{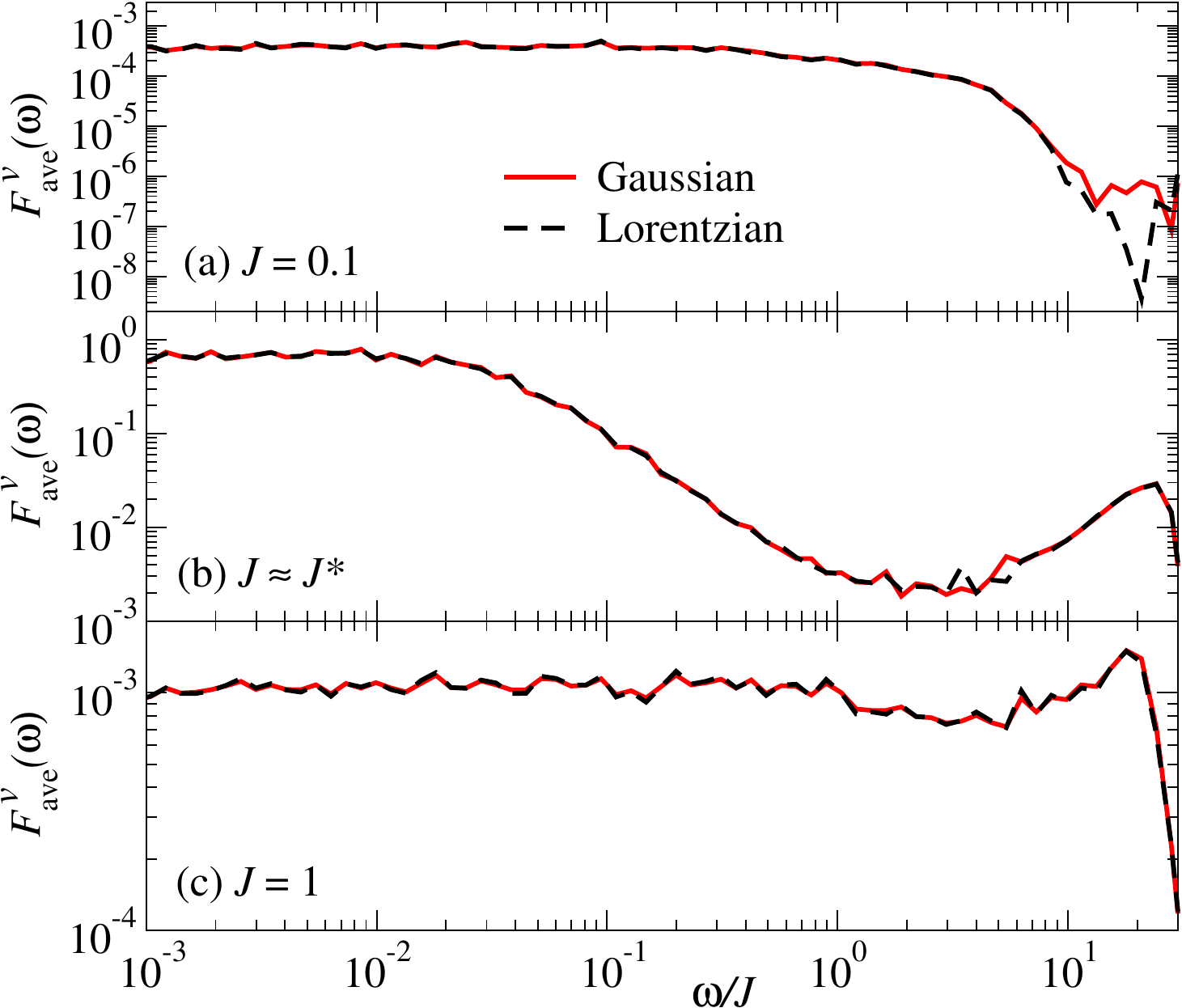}
    \vspace{-0.1cm}
    \caption{The spectral function $F^{v}_\text{ave}$ vs $\omega/J$ for (a) $J\!=\!0.1$, (b) $J\!=\!0.3$, and (c) $J\!=\!1$ obtained using Gaussian and Lorentzian broadenings. We show results for $L=22$ computed as a weighted average over the two $Z_{2}$ sectors and the quasimomentum sector with $k=6\pi/11$.}
    \label{fig:Lorentzian_spectf}
\end{figure}

For the results reported in the main text, we regularize the $\delta$ function using the Gaussian function $\exp(-\frac{x^2}{2\eta^2})\!/\!\left(\sqrt{2\pi}\eta\right)$, where $\eta\!=\!\omega_{\rm min}$ denotes a cutoff frequency with $\omega_{\rm min}\!=\!\min_n(E_{n+1}\!-\!E_{n})$. Those results are insensitive to the specific regularization used. Another common regularization is that provided by the Lorentzian $\delta(x)\approx \eta/[\pi(x^2+\eta^2)]$. In Fig.~\ref{fig:Lorentzian_spectf}, we compare the results for the spectral function reported in Fig.~\ref{fig:chi_spectf}(b) to those obtained using the Lorentzian for $L=22$. For the frequencies shown in Fig.~\ref{fig:chi_spectf}(b), the results from both regularizations agree with each other. Differences between them only emerge at higher frequencies when there is little spectral weight. In those instances, the tails of the function used do affect the results, as seen in Fig.~\ref{fig:Lorentzian_spectf}(a).\\

\section{Spectral function of the 2D TFIM\label{sec_S4}}

\begin{figure}[!h]
    \includegraphics[width=.985\columnwidth]{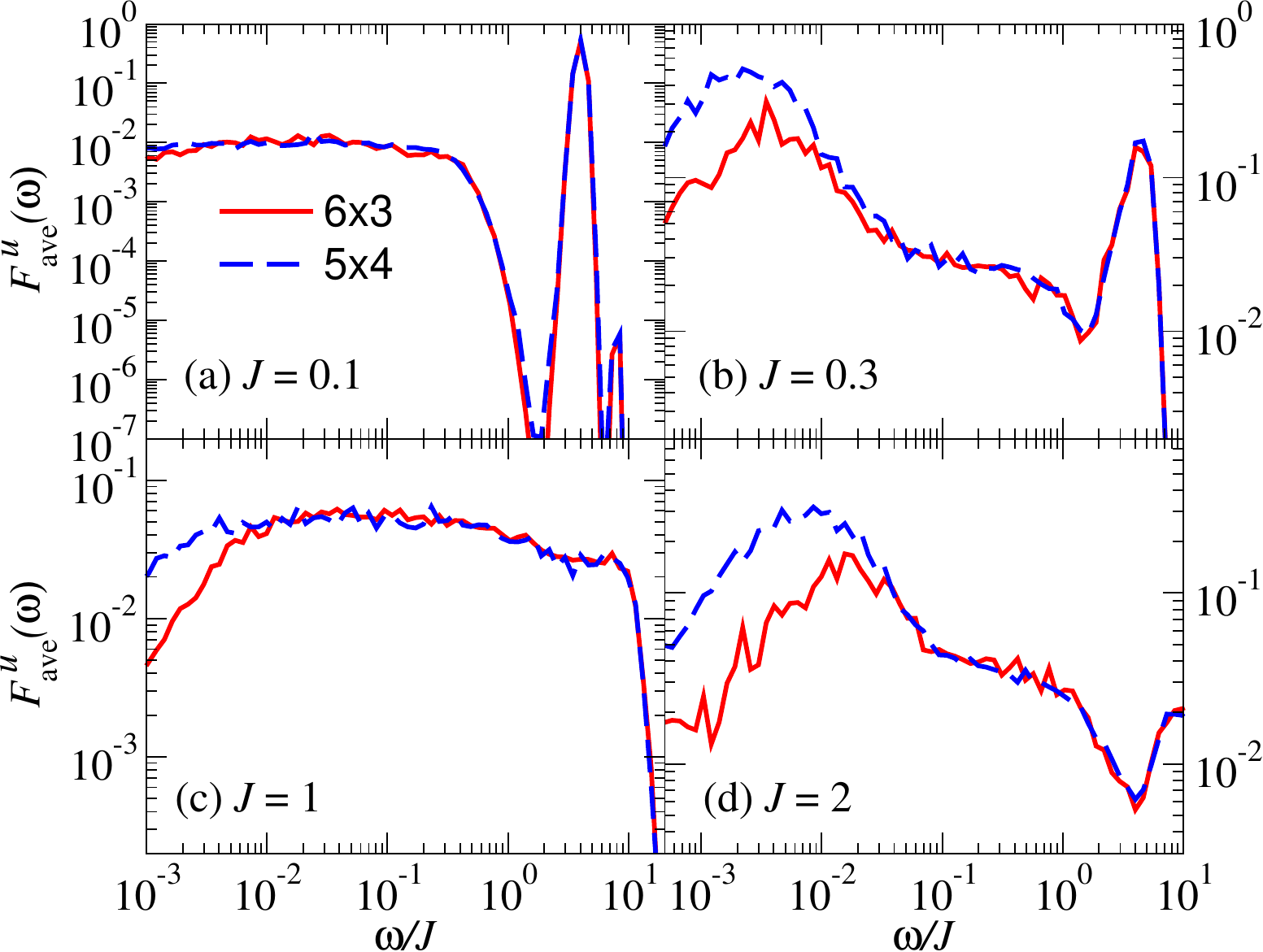}
    \vspace{-0.1cm}
    \caption{The spectral function $F^{u}_\text{ave}$ vs $\omega/J$ for (a) $J\!=\!0.1$, (b) $J\!=\!0.3$, (c) $J\!=\!1$, and (d) $J\!=\!2$. We report results obtained in the quasimomentum $k=(0,0)$ sector, averaged over all states in the $Z_{2}$, $M_{x}$, and $M_{y}$ subsectors ($\hat M_{x}$ and $\hat M_{y}$ stand for mirror symmetry in $x$ and $y$, respectively) in lattices with $L_x=6,\,L_y=3$ and $L_x=5,\,L_y=4$.}
    \label{fig:spectf_2D}
\end{figure}

In Fig.~\ref{fig:spectf_2D}, we plot the spectral function (obtained using the Gaussian broadening), for the same observable $u$ for which we computed the susceptibility reported in Fig.~\ref{fig:2D_TFIM} in the main text. We show the spectral function in four different regimes; (a) the chaotic but not ergodic regime $J\!=\!0.1$, (b) the first crossover $J\!=\!0.3$, (c) the ergodic regime $J\!=\!1$, and (d) the second crossover $J\!=\!2$. As expected, the spectral functions in both (a) and (c) exhibit a plateau at small frequencies, typical for chaotic regimes. At the two crossovers, where the susceptibility exhibits a maximum, the spectral function diverges at small frequencies as evident from panels (b) and (d), signaling slow dynamics around these crossovers.

\end{document}